\documentclass[a4paper,11pt]{article}
\usepackage{jheppub}
\usepackage{slashed}
\usepackage{booktabs}
\usepackage{graphicx}
\usepackage[utf8]{inputenc}
\usepackage[%
separate-uncertainty=true,%
]{siunitx}
\usepackage[dvipsnames]{xcolor}
\usepackage{bbm}

\usepackage[utf8]{inputenc}
\usepackage{dcolumn}
\usepackage{bm}
\usepackage{amsmath}
\usepackage{slashed}
\usepackage{float}
\usepackage{physics}

\newcommand{\diag}{\mathop{\mathrm{diag}}}

\newcommand{\CR}{\mathop{\mathrm{CR}}\nolimits}

\renewcommand{\Im}{\mathop{\mathrm{Im}}}
\newcommand{\pmns}{\text{\sc pmns}}

\newcommand{\tot}{\mathrm{tot}}
\begin{document}

\title{Heavy Neutral Leptons -- Advancing  into the PeV domain}

\author{Kevin A.\ Urqu\'ia-Calder\'on,}
\author{Inar Timiryasov,}
\author{and Oleg Ruchayskiy}

\affiliation{Niels Bohr Institute, University of Copenhagen, Blegdamsvej 17, DK-2010, Copenhagen, Denmark}
\emailAdd{kevin.urquia@nbi.ku.dk}
\emailAdd{inar.timiryasov@nbi.ku.dk}
\emailAdd{oleg.ruchayskiy@nbi.ku.dk}

\arxivnumber{2206.04540}

\abstract{Heavy neutral leptons (HNLs) are hypothetical particles able to explain neutrino oscillations and provide a mechanism for generating the baryon asymmetry of the Universe. Quantum corrections due to such particles give rise to flavor violating processes in the charged lepton sector. Based on the fact that these corrections grow with HNL masses, we improve existing constraints by orders of magnitude in mass and mixing angle. This allows us to probe part of the parameter space of leptogenesis with multi-TeV HNLs. We also show that one will be able to infer HNL parameters in a significant portion of the parameter space for TeV-PeV masses if charged lepton flavor violating signals are detected.
}

\maketitle

\section{Introduction}
Neutrinos in the Standard Model come in three different flavors.
However, unlike other fermions, neutrinos change their flavor when propagating -- a phenomenon known as \emph{neutrino oscillations} (see e.g.\ \cite[Ch.~14]{ParticleDataGroup:2020ssz}).
The Standard Model forbids such a flavor-changing transition.
Therefore neutrino oscillations imply the existence of extra quantum states, see e.g.\ \cite{deGouvea:2016qpx} for an overview of neutrino mass models.
One example of such states is given by right-chiral gauge singlet fermions, interacting with lepton doublets and the Higgs field via the Yukawa interactions.
Being gauge singlets, these states can have Majorana masses, whose scale can be arbitrary \cite{Minkowski:1977sc,Yanagida:1979as,Glashow:1979nm,Gell-Mann:1979vob,Mohapatra:1979ia,Mohapatra:1980yp,Schechter:1980gr,Schechter:1981cv}.
Compared to the Standard Model spectrum, this \emph{Type I seesaw model} contains light neutrinos, $\nu_i$, $i=1,2,3$ whose masses and mixings fit the experimental neutrino measurements \cite{Esteban:2020cvm} and, additionally, \emph{heavy neutral leptons} (HNLs) $N_I$, $I=1,2\dots$ that interact similarly to neutrinos, but are much heavier and have a suppressed interaction strength due to the mixing angle $U_\alpha^2$.\footnote{Our notations:  flavor index: $\alpha = \{e,\mu,\tau\}$; HNL masses: $M_N$; mixing angle $U_\alpha^2 = \sum_I |\Theta_{\alpha I}|^2$ where $\Theta_{\alpha I}$ is a mixing angle between flavor $\alpha$ and HNL's flavor $I$.}
HNL masses can lie anywhere from $\mathcal{O}(\SI{1}{eV})$ till $\mathcal{O}(\SI{e15}{GeV})$, with experiments limiting their interaction from above (direct experimental searches, see \cite{Agrawal:2021dbo,Abdullahi:2022jlv}) or from below (as in the case of primordial nucleosynthesis, see \emph{e.g.}
\cite{Sabti:2020yrt,Boyarsky:2020dzc,Bondarenko:2021cpc}).

HNLs can also be responsible for generating the matter-antimatter asymmetry of the Universe~\cite{Fukugita:2002hu}.
This scenario (known as \emph{leptogenesis}) has been actively developed since the 1980s (see, e.g., reviews~\cite{Davidson:2008bu,Pilaftsis:2009pk,Shaposhnikov:2009zzb,Buchmuller:2004nz,Bodeker:2020ghk}). Different mechanism of \emph{low-scale} leptogenesis have been suggested in~\cite{Akhmedov:1998qx,Asaka:2005pn,Pilaftsis:2005rv}. Recently, a unified description of these mechanisms in the case of two HNLS~\cite{Klaric:2020phc,Klaric:2021cpi} and three HNLs~\cite{Drewes:2021nqr} has been presented. In both cases, the mixing angles $\Theta_{\alpha I}$ allowed by leptogenesis could be much larger than ``naive'' seesaw expectations. This 
can be realized in a technically natural 
way~\cite{Wyler:1982dd,Leung:1983ti,Mohapatra:1986bd,Branco:1988ex,Gonzalez-Garcia:1988okv,Shaposhnikov:2006nn,Kersten:2007vk,Abada:2007ux,Gavela:2009cd,Moffat:2017feq,Drewes:2019byd} if, for example, two HNLs form a pseudo-Dirac fermion~\cite{Wolfenstein:1981kw,Petcov:1982ya}.

To mediate neutrino oscillations, HNLs should mix with neutrinos of several flavors. 
As a result,  HNLs induce charged lepton flavor violating
processes (cLFV) via tree-level \cite[see e.g.][]{Leung:1983ix,Gronau:1984ct,delAguila:2007qnc,Cvetic:2010rw,Cvetic:2012hd,Alva:2014gxa,Pascoli:2018heg,Fuks:2020att} or loop effects~\cite{Petcov:1976ff,Bilenky:1977du,Marciano:1977wx,Minkowski:1977sc,Cheng:1980tp,Lim:1981kv,Langacker:1988up,Pilaftsis:1992st,Ilakovac:1994kj,Illana:1999ww,Illana:2000ic,Pascoli:2003rq,Pascoli:2003uh,Arganda:2004bz,Gorbunov:2014ypa,Arganda:2016zvc,Abada:2015oba,Bolton:2022lrg,Bai:2022sxq}.
Non-observations of cLFV processes constrain products $|U_\alpha U_{\beta\neq \alpha}|$.
Assuming ratios between flavors, these constrains are the most powerful for $M_N \gtrsim \mathcal{O}(\SI{100}{GeV})$ not accessible via direct experimental searches \cite[see e.g.][]{deGouvea:2015euy,Fernandez-Martinez:2016lgt}, see \cite{Bernstein:2013hba,Calibbi:2017uvl} for the recent experimental status of cLFV searches.

\bigskip

In this work, we highlight the non-trivial mass dependence of the cLFV processes and demonstrate that HNLs within the TeV-PeV mass range could significantly influence the magnitude and rates of cLFV events.
More specifically, we show that for HNLs with masses $M_N \gg\SI{1}{TeV}$, the existing cLFV bounds are much stronger than previously estimated.
In fact, they are one of the \emph{the strongest} across the whole range of HNL masses.\footnote{\label{fn:1}We consider nearly degenerated HNLs, and therefore constraints from Lepton Number Violating observables, such as $0\nu\beta\beta$ effect \cite{Bolton:2019pcu} or $W^\pm W^\pm$ scattering \cite{Fuks:2020att, CMS:2022rqc} are not applicable due to approximately conserved lepton number.
The improvement stems from accounting for the one-loop diagrams with two HNLs mostly ignored previously.}

This observation suggests the feasibility of using precision frontier measurements as a strategy for investigating the elusive high-mass domain of HNLs. We show that it is possible to explore a portion of the leptogenesis parameter space previously considered inaccessible.
Furthermore, we demonstrate that potential observations of cLFV signals will allow for 
a more accurate determination of HNL parameters.

\section{Basic definitions for heavy neutral leptons}
\label{sec:basic-definitions}

In this section, we collect the basic definitions related to the Type-I Seesaw Lagrangian.
While it contains only known results, they are needed for the consistency of our exposition.
\subsection{Type I seesaw Lagrangian}
\label{subsec:seesaw_formulas}
The Lagrangian of the type-I seesaw model reads:
\begin{equation}
  \mathcal{L}_{\mathrm{SM}+\mathrm{HNL}}= \mathcal{L}_\mathrm{SM}
  + i \bar{\nu}_{\mathrm{R}I} \slashed{\partial} \nu_{\mathrm{R}I}
  - F_{\alpha I}\bar{L}_\alpha \tilde{\Phi}\,\nu_{\mathrm{R}I}
  - \frac{1}{2} \bar{\nu}_{\mathrm{R}I}^c\,M_I\, \nu_{\mathrm{R}I} + \mathrm{h.c.}
  \label{eq:lagrangian}
\end{equation}
where $\mathcal{L}_{\mathrm{SM}}$ is the SM Lagrangian, $\nu_{\mathrm{R}I}$, $I=1,\dots,\mathcal{N}$ are the right-chiral components of neutrino fields -- SM gauge singlets --, $L_\alpha$ is SM lepton doublets, $\tilde{\Phi} = i \sigma_2\Phi$ where $\Phi$ is the Higgs doublet, $F_{\alpha I}$ is a Yukawa matrix, and $M_I$ is the Majorana masses of right-handed neutrinos.
Finally, sum over $I$ and $\alpha$ is assumed in~(\ref{eq:lagrangian}).
After spontaneous symmetry breaking, the model will acquire Dirac masses, $(m_D)_{\alpha I}$ mixing right and left-handed neutrino components.
In the case $|m_D| \ll M_I$, the diagonalization of the mass matrix results in three fields $\nu_i$ with light masses, $m_i$ and $\mathcal{N}$ heavy fields $N_I$ with heavy masses $M_I$ of the shape:
\begin{equation}
  (m_\nu)_{\alpha\beta} \simeq -\sum_I (m_D)_{\alpha I} \frac{1}{M_I} (m_D)_{I\beta},
  \label{eq:see-saw}
\end{equation}
where 
\begin{equation}
  \label{eq:PMNS}
  (m_\nu)_{\alpha\beta} =\sum_{i=1}^3V^{\pmns}_{\alpha i} m_i V^{\pmns}_{i\beta}\,,
\end{equation}
$V^{\pmns}$ is the PMNS matrix \cite{Petcov:2013poa}, and $\alpha = e, \mu, \tau$.
Eq.~(\ref{eq:see-saw}) is the famous seesaw mechanism \cite{Minkowski:1977sc, Yanagida:1979as, Mohapatra:1979ia, Mohapatra:1980yp, Schechter:1980gr, Schechter:1981cv}.

After diagonalization, neutrinos in the flavor basis are in a superposition of light and heavy mass states:
\begin{equation}
  \label{eq:charge_mass_basis}
  \nu_\alpha = V^{\pmns}_{\alpha i}\,\nu_i + \Theta_{\alpha I}\,N_I^c\, = \mathcal{U}_{\alpha i}\,n_i + \mathcal{U}_{\alpha I}\,n_I\,,
\end{equation}
where $\Theta$ is the left-right neutrino mixing angle, and the $3\times (3+ \mathcal{N})$ matrix $\mathcal U$ is defined as
\begin{equation}
  \label{eq:7}
  \mathcal{U} \equiv \left(V^{\pmns}, \Theta\right)
\end{equation}
Eq.~(\ref{eq:charge_mass_basis}) makes it clear that HNLs will interact in the exact same way as active neutrinos do, with the notable exception that it will be suppressed by the mixing angle $\Theta_{\alpha I}$ \emph{forming a $3\times \mathcal{N}$ matrix} with components
\begin{equation}
  \Theta_{\alpha I} = \frac{(m_D)_{\alpha I}}{M_I}\,.
  \label{eq:theta_definition}
\end{equation}
Furthermore, we can define flavor mixing angles
\begin{equation}
  \label{eq:1}
  U_\alpha^2 \equiv \sum_I |\Theta_{\alpha I}|^2\,,
\end{equation}
to which in the case of nearly degenerate HNLs the signal would be proportional.

In addition, after diagonalization interactions between neutral leptons and SM gauge bosons get modified from the usual SM ones. These interactions read as:
\begin{align}
        \label{eq:W_vertex}
        \mathcal{L}_W &= \frac{g}{\sqrt{2}}\,\mathcal{U}_{\alpha i}\,\bar{\ell}_\alpha\,\slashed{W}^-\,P_L\,n_i + \text{h.c.}\,, \\
        \label{eq:Z_vertex}
        \mathcal{L}_Z &= \frac{g}{2c_w} \bar{n}_i \slashed{Z} \left[\mathcal{C}_{ij}^{\phantom{\ast}} P_L - \mathcal{C}_{ij}^{\ast} P_R\right] n_j\,, \\
        \label{eq:phi_pm_vertex}
        \mathcal{L}_{\phi^\pm} &= \frac{g}{\sqrt{2} m_w} \mathcal{U}_{\alpha i}\,\phi^- \left[\bar{\ell}_\alpha \left(m_i P_R - m_\alpha P_L\right) n_i \right] + \text{h.c.}\,, \\
        \label{eq:phi_0_vertex}
        \mathcal{L}_{\phi^0} &= \frac{ig}{4 m_w}\,\phi^0\,\bar{n}_i \left[\mathcal{C}^{\phantom{\ast}}_{ij} \left(m_j\,P_R - m_i\,P_L \right) + \mathcal{C}_{ij}^{\ast}\left(m_i P_R - m_j P_L\right)\right] n_j \,, \\
        \label{eq:higgs_vertex}
        \mathcal{L}_h &= -\frac{g}{4 m_w}\,h\,\bar{n}_i \left[\mathcal{C}^{\phantom{\ast}}_{ij} \left(m_j\,P_R + m_i\,P_L \right) + \mathcal{C}_{ij}^{\ast}\left(m_i P_R + m_j P_L\right)\right] n_j \,,
	\end{align}
where $\mathcal{C}_{ij} = \sum_{\alpha} \mathcal{U}_{\alpha i}^{\ast}\,\mathcal{U}_{\alpha j}^{\phantom{\ast}}$. We properly used the Feynman rules regarding Majorana particles \cite{Denner:1992vza}, especially with regards to vertices involving lepton number violating (LNV) processes or in interactions with neutral bosons that involve two Majorana particles.

There is a number of relations that the $\mathcal{U}, \mathcal{C}$ and mass matrix of neutral fermions follow that can be found in Appendix~\ref{app:unitarity_seesaw_relations}.

\subsection{Parametrization of the mixing angles} \label{sec:parametrization_mixing_angles}
A matrix $\Theta$ that automatically satisfies the seesaw relation \eqref{eq:see-saw} is given by the \textit{Casas-Ibarra parametrization} \cite{Casas:2001sr}:
\begin{equation}
  \Theta = i\,V^\pmns \sqrt{m_\nu}\,R\,\frac{1}{\sqrt{M_N}},
\end{equation}
where $m_\nu = \diag(m_1, m_2,m_3)$, $M_N = \diag(M_1,\dots, M_{\mathcal{N}})$ and the $3 \times \mathcal{N}$ matrix $R$ obeys the relation: $R^T R = \mathbbm{1}$.

The model has introduced a total of new $7\mathcal{N} -3$ different parameters to the SM. Neutrino oscillation data already constraints some of them: two different mass splittings and three mixing angles from the PMNS matrix \cite{Esteban:2020cvm}. In the active neutrino sector, we additionally have the lightest neutrino mass, two Majorana CP phases, $\delta_{\mathrm{CP}}$, one Dirac CP phase, and the discrete choice of mass hierarchy.  

Throughout this work, we are dealing with the simplest realistic case of $\mathcal{N} =2$. In this case, the lightest neutrino mass is set automatically to zero, and only a linear combination of two Majorana phases contributes, which we will call $\eta$. This leaves us with three unknown parameters in the active neutrino sector, with the already mentioned unbeknownst mass hierarchy. In this model, the $R$ matrix depends on the choice of hierarchy \cite{Ibarra:2003up, Petcov:2005jh}:

\begin{equation}
  \label{eq:casas_ibarra_matrix}
  R^{\mathrm{NO}} = \begin{pmatrix}
    0 & 0 \\
    \cos\omega & \sin\omega \\ 
    -\sin\omega & \cos\omega
  \end{pmatrix}, \hspace{1cm} 
  R^{\mathrm{IO}} = \begin{pmatrix}
    \cos\omega & \sin\omega \\ 
    -\sin\omega & \cos\omega \\
    0 & 0
  \end{pmatrix},\qquad \omega \in \mathbbm{C}\,.
\end{equation}  
It is useful to define the sum of the absolute value squared of all the elements of $|\Theta_{\alpha I}|^2$, $U_{\mathrm{tot}}^2$, in terms of Casas-Ibarra parameters, and HNL and active neutrino masses.
The expression becomes simple when considering two HNLs with equal masses:
\begin{equation}
  \label{eq:Utot_CI}
  U_{\tot}^2 = \sum_{\alpha I} \left|\Theta_{\alpha I}\right|^2 = \frac{\sum_i m_i}{M}\,\cosh(2 \Im{\omega})\,,
\end{equation}
the reader may find a more general expression in \cite{Eijima:2018qke} for non-degenerate HNL masses. 

\section{Decoupling behaviour}
cLFV observables receive contributions from HNLs, appearing in the loop once
or twice: compare the diagram (a) with the diagrams (b) and (c) in Figure~\ref{fig:mu_e_conversion} (see Appendix~\ref{app:pheno_cLFV} for details).
Two-HNL contributions bear similarity with the ``penguin diagrams'' in flavor physics~\cite[see e.g.,][]{Shifman:1995hc} and appear whenever $Z$ or Higgs bosons contribute to the diagrams.
Schematically, the conversion rate $\mu\to e$ on a nucleus (the process that will give the strongest bounds) has the form   
\begin{equation}
  \label{eq:Br}
  \CR(\mu\to e)\propto \Lambda_{e\mu}U_{\tot}^4 \Bigl|c_0(M_N) + U_{\tot}^2 c_2(M_N)\Bigr|^2
\end{equation}
where $U_{\tot}^2 = \sum_{\alpha,I} \bigl|\Theta_{\alpha I}\bigr|^2$ and
\begin{equation}
  \label{eq:3}
  \Lambda_{\alpha\beta} \equiv \frac{\left| \sum_{I} \Theta_{\alpha I} \Theta_{\beta I}^*\right|^2}{U_\tot^4}
\end{equation}
The functions
$c_i(M_N)$ scale for HNL mass $M_N \gg m_W$ as:
\begin{equation}
  \label{eq:2}
  c_2(M_N) \propto\frac{M_N^2}{m_W^2} \gg c_0(M_N) \propto \log\left(\frac{M_N}{m_W}\right)
\end{equation}
As a result, although subdominant due to $U_\tot^2\ll 1$, the strong mass dependence of $c_2(M_N)$ makes the second term larger than the first one.
As a result, HNL contributions to cLFV processes become much stronger at large masses than previously estimated.\footnote{The process $\ell_\alpha \to 3\ell_\beta$ has branching ratio of the form, similar to~(\ref{eq:Br}). The process $\ell_\alpha \to \ell_\beta \gamma$ does not receive contributions from ``penguins''.}

The above result seemingly contradicts the ``Decoupling Theorem'' \cite{Appelquist:1974tg}, as loop effects of heavy particles grow with mass.
However, theories that undergo spontaneous symmetry breaking violate this theorem if dimensionless couplings -- in our case, the Yukawa coupling that determines the Dirac mass  --- are sent to infinity (see e.g.\ \cite{Collins:1978wz, Cheng:1991dy, Tommasini:1995ii, DHoker:1984mif, DHoker:1984izu}, or the book \cite[Chapter~8]{Collins:1984xc}).
At a technical level, this happens because the coupling between HNLs and Goldstone bosons is proportional to the Yukawa couplings, and in turn, to the Dirac Mass term $m_D \sim \Theta M_N$ (see Eq.~\ref{eq:theta_definition}).
If one considers $\Theta$ and $M_N$ as independent parameters, then Dirac mass term grows  as $M_N\to \infty$, and there is no decoupling.
On the other hand, keeping Yukawa constants fixed implies $ |\Theta| \propto M_N^{-1}$.
As a result, the effect~(\ref{eq:Br}) disappears.

By increasing $M_N$ at fixed
  $|\Theta|$ the HNL decay width $\Gamma_N \propto |\Theta|^2
  M_N^3$ exceeds $\frac 12M_N$ (equivalently, Yukawa coupling constant
  of neutrino exceeds $\sqrt{4\pi}$).
  The corresponding \emph{non-perturbative region} is shaded in gray
  below. 
  See \cite{Chanowitz:1978mv, Durand:1989zs, Fajfer:1998px, Ipek:2018sai}, and Appendix~\ref{app:higher_order_terms} for a  brief discussion on the stability on second-order terms.

Indeed, the non-decoupling behaviour of HNLs implies that when we match the amplitudes to a series of EFT operators, we will find that such operators will depend on HNL masses or on their Yukawa couplings, and this is exactly what happens when one does a one-loop matching to different EFT operators \cite{Zhang:2021jdf}.

\begin{figure}[!t]
  \centering
  \includegraphics[width=\linewidth]{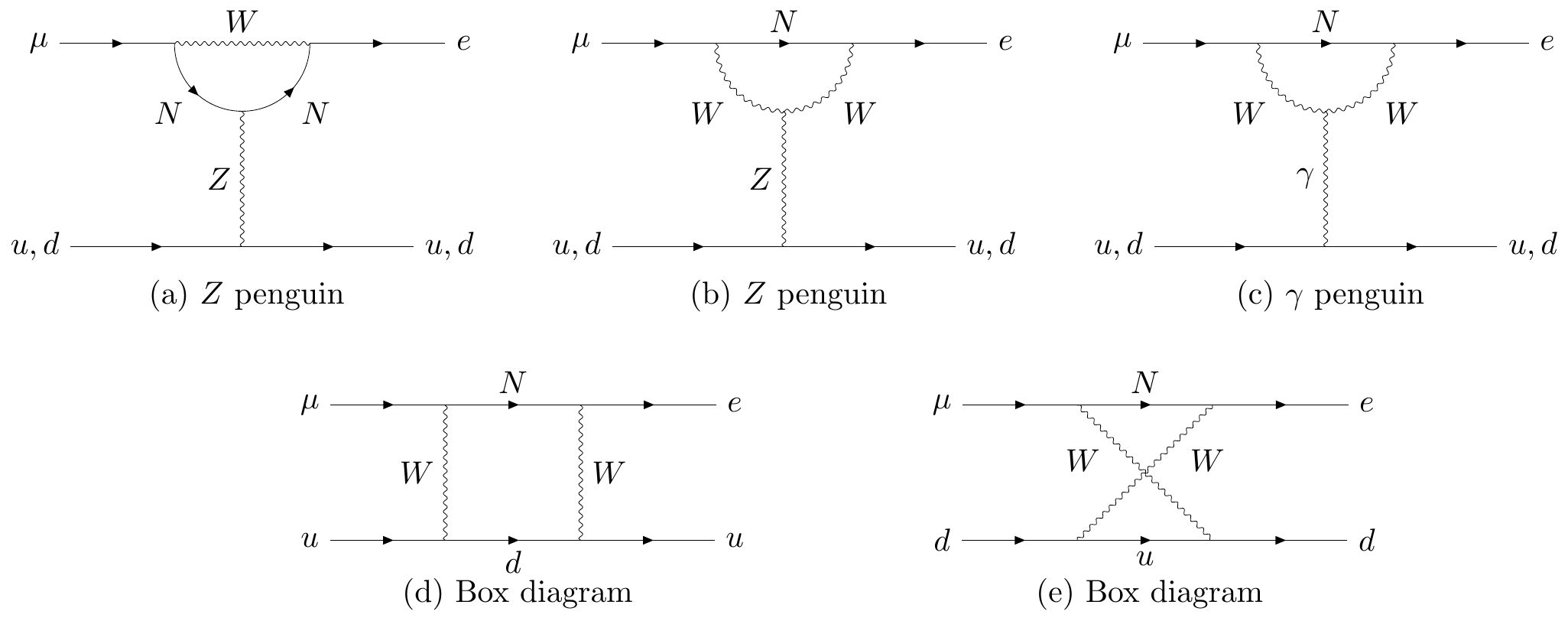}
  \caption[cLFV diagrams with 1 and 2 HNLs]{One-loop diagrams contributing to
    the $\mu\to e$ conversion process, including 2 HNL ``penguin'' diagrams
    (a). Diagrams containing HNLs that dress external lines are not shown (see Appendix~\ref{app:diagrams} for the full list of such diagrams).}
  \label{fig:mu_e_conversion}
\end{figure}

\begin{figure}[!t]
  \centering
  \resizebox{\linewidth}{!}{\includegraphics[height=1cm]{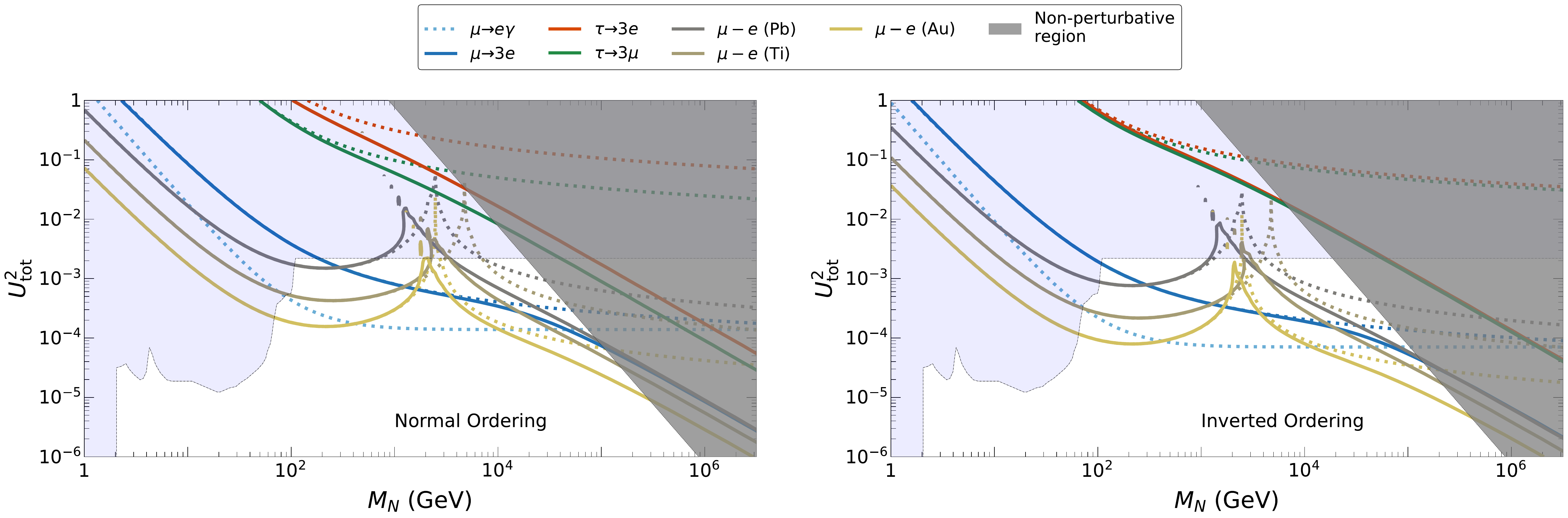}}
  \caption[Improved HNL limits.]{Improved experimental limits on HNLs coupling from charged lepton flavor violation processes (solid lines) for the normal ordering (left) and inverted ordering (right).
    The strongest bound comes from the $\mu\to e$ conversion in gold~\cite{SINDRUMII:2006dvw}.
    Blue shaded region: previous constraints, not including cLFV.
    Gray shaded region: non-perturbative regime (see text for details).
    Dotted lines: previous cLFV constrains from the works \cite{Alonso:2012ji,deGouvea:2015euy,Abada:2015oba,Fernandez-Martinez:2016lgt,Abada:2018nio}.
    Results are expressed in terms of $U_\tot^2$ for flavor ratios~(\protect\ref{eq:4}), see text for details.
  }
  \label{fig:bound_improvements}
\end{figure}

\section{Results}

Using the proper expression for the cLFV branching ratios/conversion rates, we reinterpret the existing limits on HNL parameters as shown in Figure~\ref{fig:bound_improvements} (solid lines).
To efficiently present and compare our results, we fix ratios of $U_\alpha^2$ to the benchmark values, dictated by neutrino oscillations \cite{Agrawal:2021dbo}:
\begin{equation}
  \label{eq:4}
  U_e^2 : U_\mu^2 : U_\tau^2  = \left\{
      \begin{array}{ll}
        0.06  : 0.48 : 0.46 & \text{(NO)} \\
        0.33 : 0.33 : 0.33  & \text{(IO)}
      \end{array}
      \right.
\end{equation}
(see \cite{Tastet:2021vwp}) and express all the limits in terms of the total mixing angle, $U_\tot^2$.
Compared to the electroweak precision data constraints~\cite{Fernandez-Martinez:2016lgt}, our bounds improve by as much as \emph{three orders of magnitude}.
Compared to the previous cLFV constrains (not taking into account the $c_2(M_N)$ terms), we improve the bounds by roughly \emph{one order of magnitude} for masses $\mathcal{O}(\SI{100}{TeV})$.
To the best of our knowledge, the constraints from $\mu\to e$ conversion in gold \cite{SINDRUMII:2006dvw} have not been previously used to derive bounds on HNL parameters in the multi-TeV region, and the only work that took into account $c_2(M_N)$ terms when deriving the cLFV limits was~\cite{deGouvea:2015euy} that however did not explore them beyond $\sim\SI{10}{TeV}$.

The maximal mass, probed by these cLFV processes, for degenerate HNLs, is extended from $\sim \SI{e5}{GeV}$ to about $\SI{e6}{GeV}$.
Our limits are stronger than LHC searches in the range $M_N \gtrsim \SI{20}{GeV}$ \cite{CMS:2018iaf,ATLAS:2019kpx,CMS:2022fut,ATLAS:2022atq}.
Future cLFV experiments \cite{Mihara:2013zna} may probe HNLs up to incredible masses of $M_N \sim \SI{3e7}{GeV}$ (\emph{i.e.}\
$\SI{30}{PeV}$). 
Our results get more pronounced as Yukawa couplings grow.

The constraints from $\mu\to 3e$ and muon conversion in the case of HNLs with a hierarchical mass spectrum are presented Appendix~\ref{app:pheno_cLFV}.

\begin{figure}[!t]
  \centering \resizebox{\linewidth}{!}{\includegraphics[height=1cm]{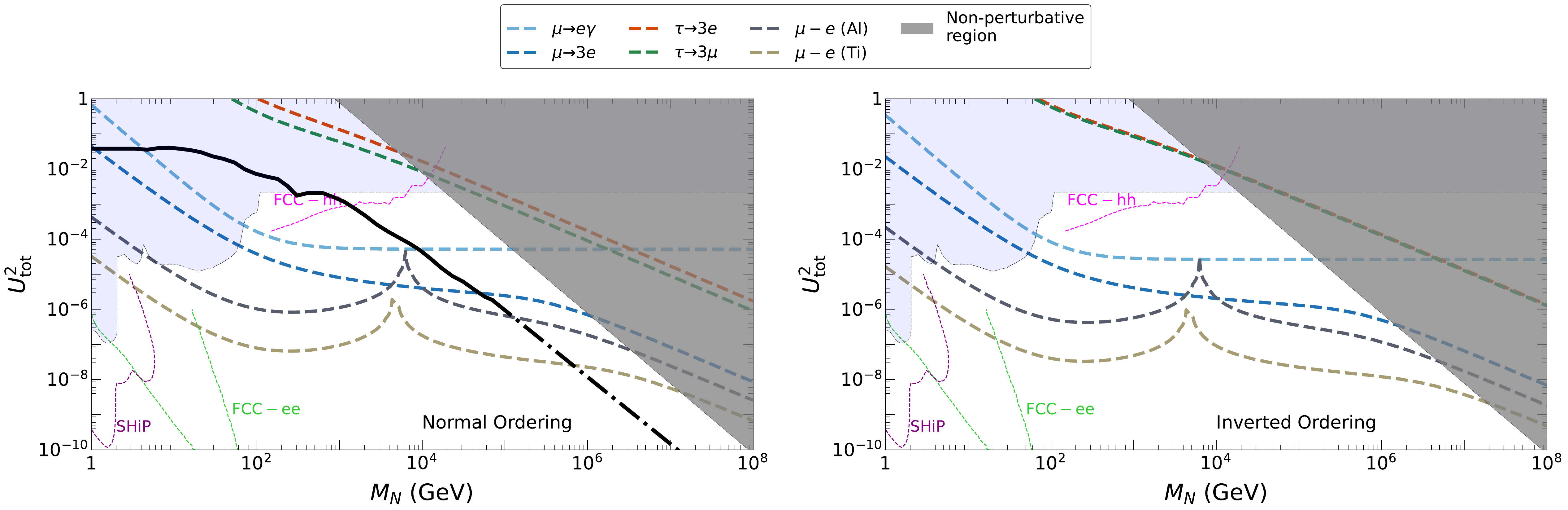}} \caption{Constrains on HNL parameters from future cLFV experiments for the normal Ordering (left) and inverted Ordering (right).
    Two HNL contributions are taken into account where applicable.
    Notice the different range of y-axes as compared to Figure~\protect\ref{fig:bound_improvements}.
    Maximal probed mass may reach $\SI{3e7}{GeV}$ should the recently proposed $\mu \to e$ conversion experiment  \protect\cite{CGroup:2022tli} be performed.
    Results are expressed in terms of $U_\tot^2$ for flavor ratios~(\protect\ref{eq:4}), see text for details.
    The black line in the right plot is the upper boundary of the parameter space where successful leptogenesis with 3 HNLs is possible \protect\cite{Drewes:2021nqr} (dashed-dotted part is an extrapolation to guide the eye).
    The sensitivity reach of several future experiments (SHiP \cite{SHiP:2018xqw}, FCC-ee \cite{Blondel:2014bra}, FCC-hh \cite{Pascoli:2018heg}) is shown as thin dashed lines.}
  \label{fig:future_sensitivity}
\end{figure}

\begin{figure}[!h]
  \centering
    \includegraphics[width=\linewidth]{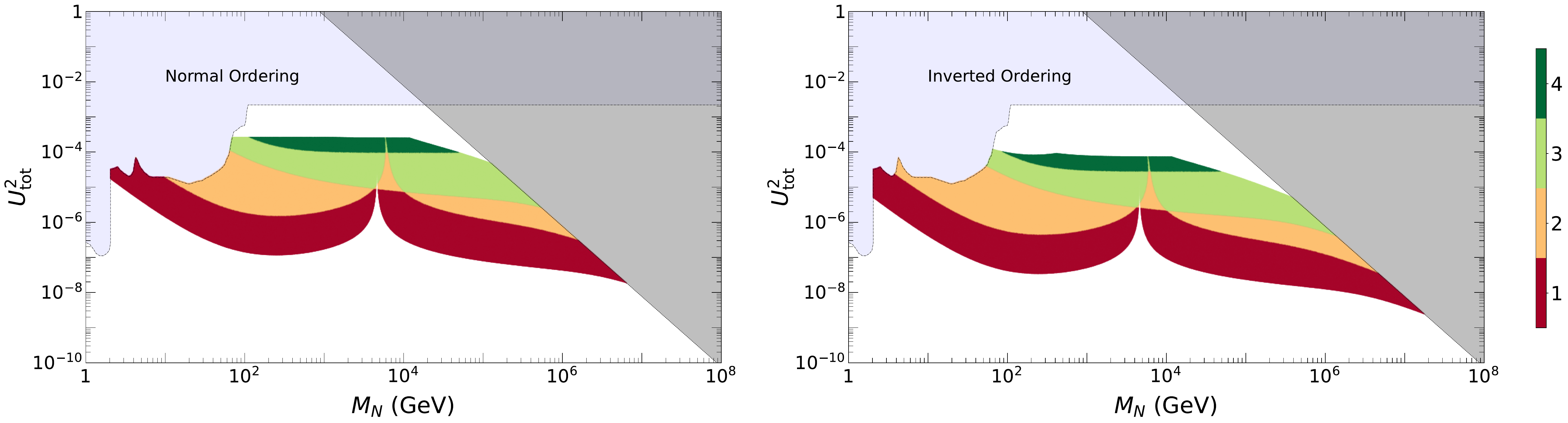}
  \caption[Parts of the parameter space where the discovery of the HNLs are possible with future cLFV experiments.]{Parameter space probed by future cLFV experiments for the normal ordering (left) and inverted ordering (right).
    Color bands illustrate where one or several future cLFV experiments have sensitivity and thus can observe a signal.
    Observation of a signal in two or more experiments  (orange to green bands) will allow for
    reconstructing parameters of realistic HNL models.
 The white region is explored by the current cLFV experiments.}
  \label{fig:discovery_zone}
\end{figure}
\begin{figure}[!h]
  \centering
  \includegraphics[width=0.8\linewidth]{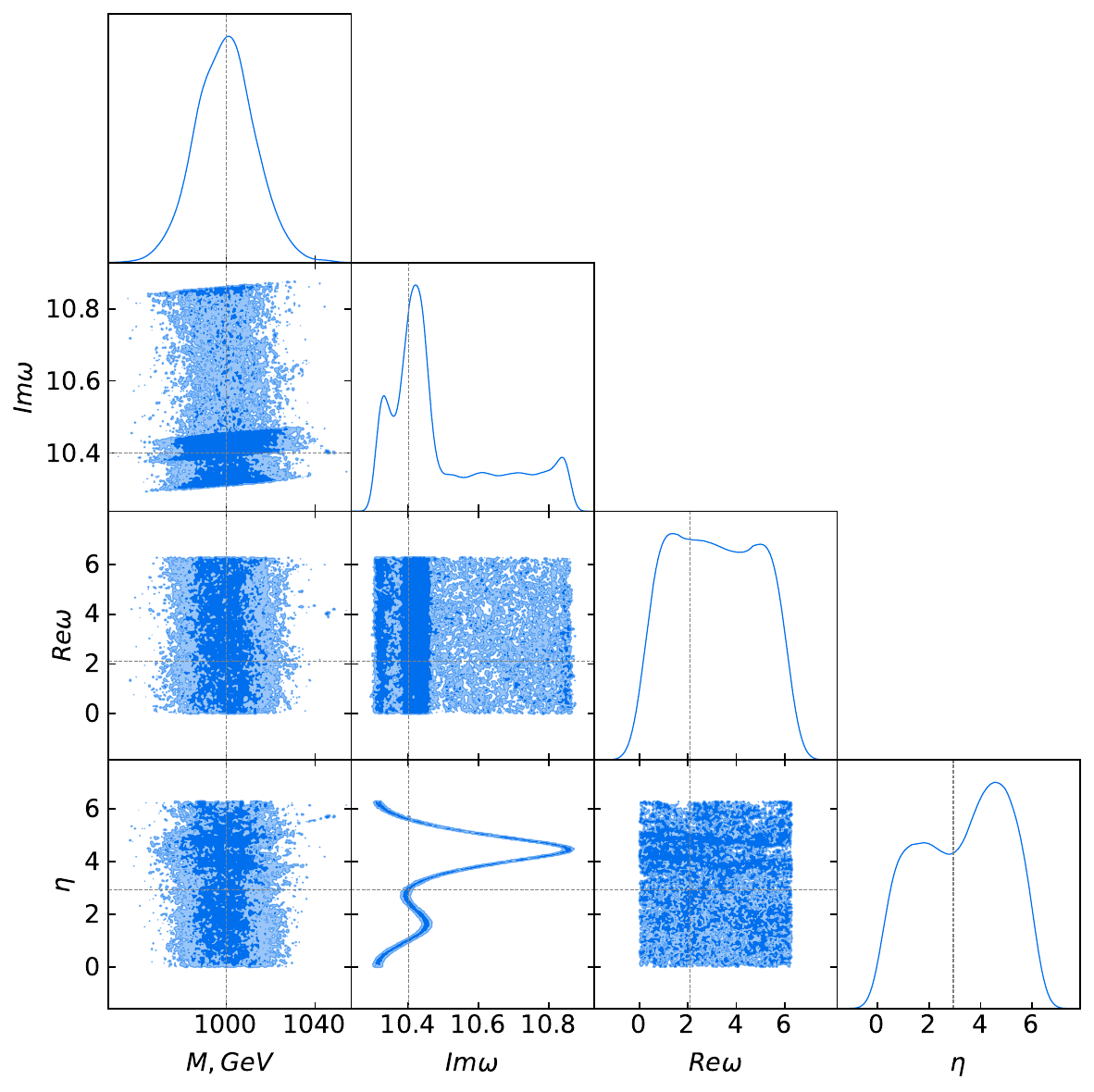}
  \caption[Mock reconstruction of HNL parameters]{Reconstruction of HNL parameters in case of two cLFV detections.
    HNL parameters, explaining neutrino data \cite{Esteban:2020cvm}, assuming normal ordering, and that the provided cLFV signals are within the future discovery zone (\textit{c.f.} Figure~\ref{fig:discovery_zone}).
    We built a likelihood function, depending on four parameters (shown in the plots as straight lines) and explored it with MCMC sampling. $\omega$ is the mixing angle from the Casas-Ibarra matrix and $\eta$ is the Majorana Phase (see Sec.~\ref{sec:parametrization_mixing_angles}). 
    Dotted grey lines show the ``true'' parameters.
  }
  \label{fig:reconstruction}
\end{figure}

\section{Indirect detection of TeV--PeV HNLs}

In the wide range of HNL parameters, two or more future cLFV experiments can detect a signal, as Figure~\ref{fig:discovery_zone} demonstrates.
Non-trivial mass dependences of $\mu\to e$ conversion rates and $\mu\to 3e$ processes open a possibility to recover HNL parameters in this case.
Similar idea has been explored in the past, \cite[see~e.g.][]{Chu:2011jg,Alonso:2012ji,Hambye:2013jsa}, albeit without ``penguin'' contributions that change the mass dependence.

Consider the simplest HNL model explaining neutrino oscillation data
-- two HNLs with almost degenerate masses.\footnote{To avoid confusion we note
  that two
  nearly degenerate HNLs can explain neutrino masses in
  the Type-I Seesaw considered here. In the case of Inverse
  \cite{Mohapatra:1986aw, Mohapatra:1986bd, Bernabeu:1987gr} or Linear
  \cite{Akhmedov:1995ip, Akhmedov:1995vm} Seesaw models \emph{at least four}
   HNLs with opposite CP phases are needed.}  The cLFV branching ratios~(\ref{eq:Br}) depend in this case on three parameters: common mass, $M_N$, total mixing angle, $U_{\tot}^2$, and $\Lambda_{e\mu}$.
While flavor mixing angles, $U_\alpha^2$ can still vary in the wide range,
$\Lambda_{e\mu}$ changes only weakly if the mass ordering and the CP phase is
known~\cite[see e.g.][]{Patterson:2015xja,DUNE:2021mtg}.
As a result, given \emph{two} cLFV measurements ($\mu\to e$ conversions and $\mu\to 3e$ measurement, Table~\ref{tab:branching_coefficients}), the HNL parameters can be reconstructed.

To demonstrate the feasibility of parameter reconstruction, we have randomly assumed HNL parameters in the light green region of Figure~\ref{fig:discovery_zone} and computed the rates of the cLFV processes. 
Assuming that these were the measured rates (with 50\% measurement error), we defined the log-likelihood as a mean square deviation between the predicted and ``measured'' rates. We performed an MCMC scan of the parameter space using the \texttt{emcee} pacakge~\cite{Foreman-Mackey:2012any}.
Using Kernel Density Estimation method implemented in \texttt{GetDist}~\cite{Lewis:2019xzd},
we successfully recovered both mass and the total mixing angle, see Figure~\ref{fig:reconstruction}.

\begin{table}[!t]
  \centering
  \begin{tabular}{l|l|l}
    \toprule
    Process & Current bounds & Future limits \\
    \midrule
    \multicolumn{3}{c}{Processes, proportional to $\Lambda_{e\mu}$} \\
    \midrule
    $\mu\to e\gamma$ & $4.2\times 10^{-13}$ \cite{MEG:2016leq} & $6\times 10^{-14}$ \cite{MEGII:2018kmf} \\
    $\mu\to 3e$ & $1.0\times 10^{-12}$ \cite{SINDRUM:1987nra} & $1 \times 10^{-16}$ \cite{Blondel:2013ia,Berger:2014vba} \\
    \midrule
    $\mu\to e$ & $4.6\times 10^{-11}$ (Pb) \cite{SINDRUMII:1993gxf} & $1\times 10^{-17}$ (Al) \cite{Mu2e:2014fns,COMET:2018auw} \\
    conversion & $4.3 \times 10^{-12}$ (Ti) \cite{SINDRUMII:1996fti} & $1 \times 10^{-19}$ (Ti) \cite{CGroup:2022tli} \\
            & $7.0 \times 10^{-13}$ (Au) \cite{SINDRUMII:2006dvw}
                             &  \\
    \midrule
    $Z\to e\mu$ & $7.5 \times 10^{-7}$ \cite{ATLAS:2014vur} & $1\times 10^{-10}$ \cite{Dam:2018rfz} \\
    $h\to e\mu$ & $6.1\times 10^{-5}$ \cite{ATLAS:2019old} &  \\
    \midrule
    \multicolumn{3}{c}{Processes, proportional to $\Lambda_{e\tau}$} \\
    \midrule
    $\tau\to e\gamma$ & $3.3\times 10^{-8}$ \cite{BaBar:2009hkt} & $3\times 10^{-9}$ \cite{Belle-II:2018jsg} \\
    $\tau\to3e$ & $2.7\times 10^{-8}$ \cite{Hayasaka:2010np} & $5\times 10^{-10}$ \cite{Belle-II:2018jsg} \\
    $Z\to e\tau$ & $9.8\times 10^{-6}$ \cite{OPAL:1995grn} & $1\times 10^{-9}$ \cite{Dam:2018rfz} \\
    $h\to e\tau$ & $4.7 \times 10^{-3}$ \cite{ATLAS:2019pmk} &  \\
    \midrule
    \multicolumn{3}{c}{Processes, proportional to $\Lambda_{\mu\tau}$} \\
    \midrule
    $\tau\to\mu\gamma$ & $4.4\times 10^{-8}$ \cite{BaBar:2009hkt} & $1\times 10^{-9}$ \cite{Belle-II:2018jsg} \\
    $\tau\to3\mu$ & $2.1\times 10^{-8}$ \cite{Hayasaka:2010np} & $5 \times 10^{-10}$ \cite{Belle-II:2018jsg} \\
    $Z\to \tau\mu$ & $1.2\times 10^{-5}$ \cite{DELPHI:1996iox} & $1 \times 10^{-9}$ \cite{Dam:2018rfz} \\
    $h\to \tau\mu$ & $2.5\times 10^{-3}$ \cite{CMS:2017con} &  \\
    \bottomrule
  \end{tabular}
  \caption[Present and future cLFV measurements]{Present cLFV limits and
    future projections for various $\Lambda_{\alpha\beta}$ processes.}
  \label{tab:branching_coefficients}
\end{table}

\section{Discussion}
Heavy neutral leptons couple to neutrinos of different flavors and thus give rise to charged lepton flavor violating processes.
Many laboratories worldwide  (Table~\ref{tab:branching_coefficients}) pursue searches for these cLFV processes.
Their negative results provide constraints on HNL parameters, see e.g.\
\cite{Fernandez-Martinez:2016lgt}.
For $\mu\to e$ conversion rate, the HNL contributions to cLFV processes contain terms proportional to the $U_\tot^4$ and those, proportional to the $U_\tot^8$ (with the common factor $\Lambda_{e\mu}$, Eq.~\eqref{eq:3}).
The latter terms have been neglected in the past.
However, for $M_N \gg m_W$, the term $U_\tot^8$ is multiplied by the growing function in mass, $(M_N/m_W)^4$.
Therefore, for $M_N \gg m_W$ this second term becomes dominant, making constraints stronger by orders of magnitude, as we showed in this work (see Figure~\ref{fig:bound_improvements} for current and Figure~\ref{fig:future_sensitivity} for future sensitivity).
The resulting constraints for realistic HNL models (\textit{i.e.}
those where HNLs mediate neutrino oscillations) allow us to reach limits on the mixing angle compared to the best direct searches for masses up to $\sim\SI{30}{PeV}$.
In particular, these constraints exceed by orders of magnitude in mass reach and coupling sensitivity those that can be achieved with direct searches for HNLs at FCC-hh \cite{Alva:2014gxa,Pascoli:2018heg,Fuks:2020att,Abdullahi:2022jlv}.

We derived our main results for the case of the 2 HNL model, but they are readily generalizable to the models with 3 HNLs (see Appendix~\ref{app:pheno_cLFV}). In this case, cLFV measurements can probe part of the parameter space of the leptogenesis \cite{Drewes:2021nqr} as Figure~\ref{fig:future_sensitivity} demonstrates. See also \cite{Granelli:2022eru}.

The present analysis has assumed benchmark flavor ratios~(\ref{eq:4}).
Scanning over the whole allowed range of $\Lambda_{\alpha\beta}$ does not qualitatively change our results (see also \cite{Abada:2021zcm}).
In particular, in models with 2 HNLs, even in the case of normal ordering, where $U_e^2 \ll U_{\mu,\tau}^2$, the processes involving LFV in the $e\mu$ sector dominate the current and future limits.
In the models with 3 HNLs the suppression of the electron mixing may become even more extreme \cite{Abdullahi:2022jlv}, potentially making $\mu\tau$ channels competitive.
This demonstrates importance of the future searches, such as TauFV~\cite{Alekhin:2015byh,Beacham:2019nyx,EuropeanStrategyforParticlePhysicsPreparatoryGroup:2019qin} at the CERN Beam Dump Facility \cite{Ahdida:2019ubf} measurement $\tau\to 3\mu$ processes at the level $\sim 10^{-10}$.

\begin{acknowledgments}
We would like to thank Serguey Petcov, Juraj Klaric, and Alessandro Granelli for informing us about their upcoming related work.
O.R.\ would like to thank M.~Shaposhnikov for useful discussions and for sharing with us the thesis~\cite{Kirk:2015msc}.
O.R.\ and I.T.\ would like to thank the Instituto de Fisica Teorica (IFT JAM-CSIC) in Madrid for support via the Centro de Excelencia Severo Ochoa Program under Grant CEX2020-001007-S, during the Extended Workshop ``Neutrino Theories'', where this work has been completed.
This project has received funding from the European Research Council (ERC) under the European Union's Horizon 2020 research and innovation programme (GA 694896) and from the Carlsberg Foundation.
I.T.\ acknowledges support from the European Union's Horizon 2020 research and innovation program under the Marie Sklodowska-Curie grant agreement No.~847523 `INTERACTIONS'.
\end{acknowledgments}

\appendix
\renewcommand{\thefigure}{\thesection.\arabic{figure}}
\section{Unitarity and Seesaw relations}
\label{app:unitarity_seesaw_relations}
The matrix elements $\mathcal{U}_{\alpha i}$ of the matrix $\mathcal{U}$, defined by Eq.~\eqref{eq:7},  obey the following unitarity and seesaw relations~\cite{Pilaftsis:1992st}:
    \begin{align}
		&\sum_{i = 1}^{3 + \mathcal{N}}\mathcal{U}_{\alpha i}^{\phantom{\ast}}\,\mathcal{U}_{\beta i}^{\ast} = \delta_{\alpha \beta}\,, \label{eq:unitarity_1} \\
		&\hspace{-0.2cm}\sum_{\alpha  = e, \mu, \tau }\hspace{-0.2cm}\mathcal{U}_{\alpha i}^{\phantom{\ast}}\,\mathcal{U}_{\alpha j}^{\ast} = \mathcal{C}_{ij}\,, \label{eq:unitarity_2} \\
		&\sum_{i = 1}^{3 + \mathcal{N}}\mathcal{U}_{\alpha i}^{\phantom{\ast}}\,\mathcal{C}_{ij}^{\phantom{\ast}} = \mathcal{U}_{\alpha i}\,, \hspace{1cm} \sum_{i = 1}^{3 + \mathcal{N}}\mathcal{C}_{ij}^{\phantom{\ast}}\,\mathcal{C}_{ik}^{\ast} = \mathcal{C}_{kj}\,, \label{eq:unitarity_3} \\
		&\sum_{i = 1}^{3 + \mathcal{N}}\mathcal{U}_{\alpha i}^{\phantom{\ast}}\,\mathcal{U}_{\beta i}^{\phantom{\ast}}\,m_i = \sum_{i = 1}^{3 + \mathcal{N}}\mathcal{U}_{\alpha i}^{\phantom{\ast}}\,\mathcal{C}_{ij}^{\ast}\,m_i = \sum_{i = 1}^{3 + \mathcal{N}}\mathcal{C}_{ij}^{\phantom{\ast}}\,\mathcal{C}_{ik}^{\phantom{\ast}}\,m_i = 0\,. \label{eq:unitarity_4}
	\end{align}
	Relation (\ref{eq:unitarity_1}) is the unitarity of the $\mathcal{U}$ matrix, whereas (\ref{eq:unitarity_2}) defines the coupling between the $Z$ boson and $n_i n_j$. The equalities (\ref{eq:unitarity_3}) are a consequence of two previous ones, and the last set of equivalences are nothing but the Seesaw relation, expressed in Eq.~(\ref{eq:see-saw}), but written in terms of $\mathcal{U}$ and $\mathcal{C}$.	

\section{Phenomenology of cLFV processes}
\label{app:pheno_cLFV}
\subsection{Relevant branching ratios and rates}
We present all the pertinent diagrams in Appendix~\ref{app:diagrams}.
We calculated the corresponding amplitudes using the \texttt{Mathematica} packages: \texttt{FeynCalc} \cite{Shtabovenko:2020gxv}, \texttt{FeynArts} \cite{Hahn:2000kx} and \texttt{Package-X} \cite{Patel:2016fam} and cross-checked the results against existing literature
\cite{Minkowski:1977sc, Marciano:1977wx, Lim:1981kv, Riazuddin:1981hz, Langacker:1988up, Cheng:1980tp, Ilakovac:1994kj, Chang:1994hz, Pilaftsis:1998pd, Ioannisian:1999cw, Deppisch:2005zm, Pilaftsis:2005rv, Deppisch:2010fr, Alonso:2012ji}. 
To include the ``HNL penguin'' diagrams, we modified the \texttt{FeynRules} model file \texttt{HeavyN} \cite{Alva:2014gxa,Degrande:2016aje,Pascoli:2018heg} and  \cite{nloFRModel} and added previously missing vertices involving the coupling between two HNLs and neutral bosons -- those in Eq.~\eqref{eq:Z_vertex}, \eqref{eq:phi_0_vertex} and \eqref{eq:phi_pm_vertex}. The updated model is available 
at \cite{repo}.

\begin{figure}[H]
    \centering
    \includegraphics[width = 0.75\linewidth]{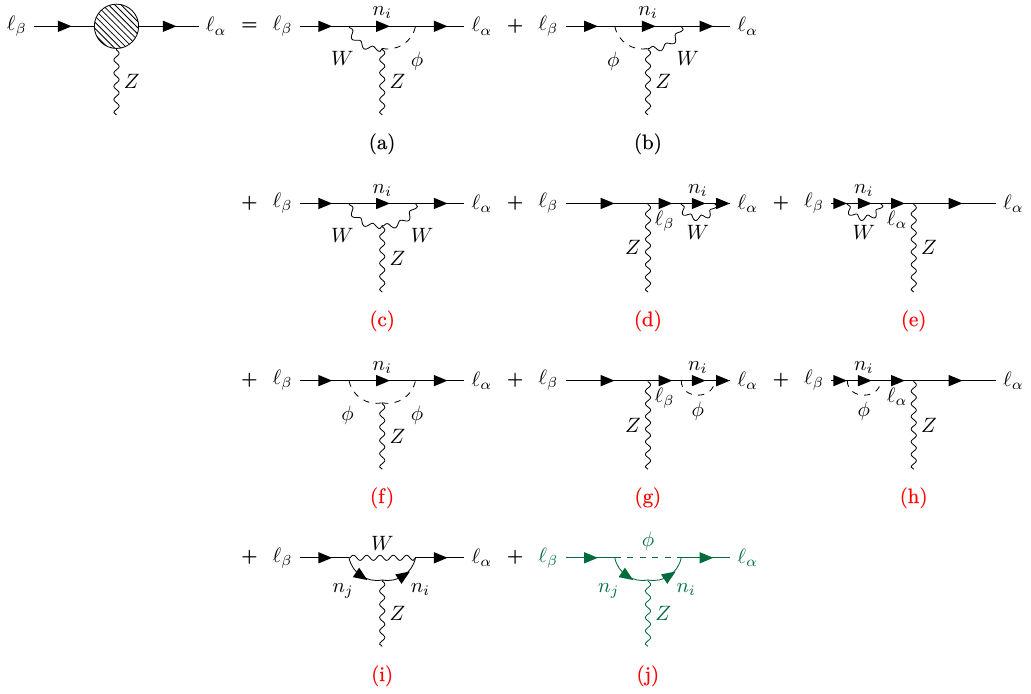}
    \caption{Diagrams contributing to the $Z \ell_\alpha \ell_\beta$ vertex. 
    The diagrams in green are new, accounting for two-HNL interactions.
    The diagrams with red labels have an UV divergence that ultimately vanishes.}
    \label{fig:z_diagrams}
\end{figure}

The diagrams that display the apropos asymptotic behavior in mass are shown below in Appendix~\ref{app:diagrams} in green. 
The anomalous $Z\ell_\alpha\ell_\beta$ coupling shown in Fig.~(\ref{fig:z_diagrams}) also gives a non-decoupling effect on both $\ell_\beta\to 3\ell_\alpha$ and muon conversion in a nucleus processes.

\interfootnotelinepenalty=10000

The amplitude of diagram (j) in Fig.~(\ref{fig:z_diagrams}) contains two different parts, one dependent on $\mathcal{C}^{\phantom{\ast}}_{ij}$ and another on $\mathcal{C}^{\ast}_{ij}$, the latter is suppressed in the case of degenerate HNLs (see \ref{app:Theta_combinations}). Both parts are asymptotic in mass, the two of them get two powers in mass from the couplings between HNLs and Goldstone bosons. The former also gets two additional powers in mass coming from the propagators of the HNLs, which cancels the mass suppression coming from the loop functions. The latter gets the momentum from the HNL propagators, but the leading terms of the loop functions are not suppressed in mass.\footnote{This can be seen from the fact that $\int \dd^4 k\,\frac{k^2}{\left(k^2 + m^2\right)^3}$ is dimensionless} Overall, both amplitudes will have a mass squared dependence.

\begin{figure}[H]
    \centering
    \includegraphics[width = \linewidth]{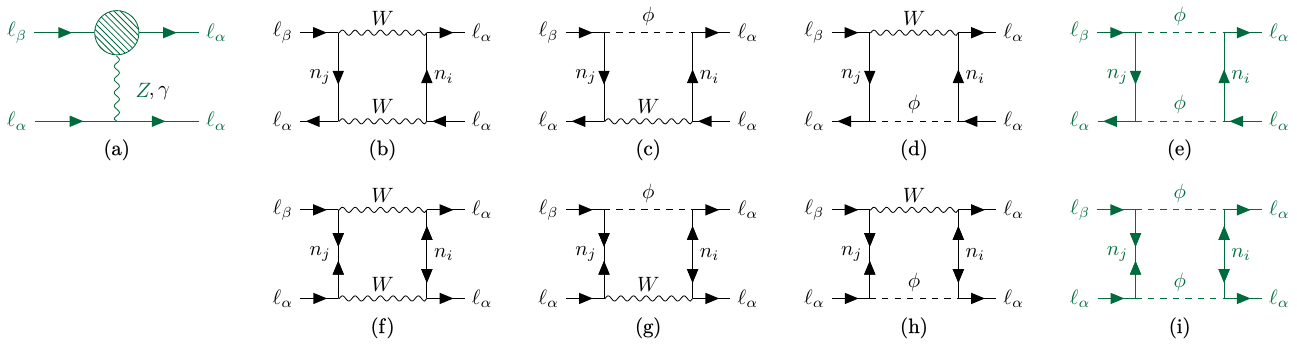}
    \caption{Diagrams contributing to the $\ell_\beta \to 3\ell_\alpha$ decay.
    The diagrams in green are new, accounting for two-HNL interactions.
     Notice that some of the two HNL contributions, coming from $Z n_i n_j$ are hidden inside the ``blob'' of the diagram (a) (given by the diagram~\ref{fig:z_diagrams}).
    }
    \label{fig:leptonic_decays}
\end{figure}

The origin of the asymptotic behavior of diagrams (e) and (i) in Fig.~(\ref{fig:leptonic_decays}) can be explain similarly. Both diagrams get four powers in mass from the couplings between the HNLs and Goldstone bosons. Diagram (e) picks up the momentum terms from the propagators of the HNLs, this gives it a loop function which removes two powers of mass from the amplitude. Whereas, diagram (i) picks up the mass terms from the propagators, but has a stronger suppression from the loop functions. All in all, both terms will have a mass squared dependence.

\begin{figure}[H]
    \centering
    \includegraphics[width = 0.75\linewidth]{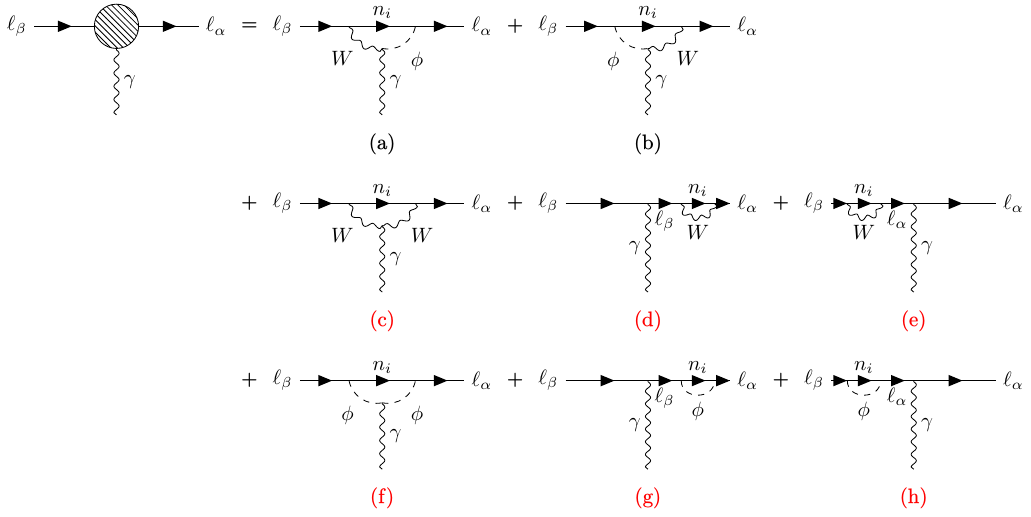}
    \caption{Diagrams contributing to the $\gamma \ell_\alpha \ell_\beta$ vertex. To this vertex there are no two-HNL contributions.
    The diagrams with red labels have an UV divergence that ultimately vanishes.}
    \label{fig:gamma_diagrams}
\end{figure}

The diagrams that have a red letter attached to them are divergent. All of these divergences vanish when we sum up the diagrams. Namely, diagrams~(c, d, e) in Figures~\ref{fig:gamma_diagrams}--\ref{fig:z_diagrams} have their divergences vanish due to the unitarity relation expressed in Eq.~({\ref{eq:unitarity_1}}), this is also the case for Diagram~(i) in Fig.~\ref{fig:z_diagrams}. Whereas for Diagram~(j) in Fig.~(\ref{fig:z_diagrams} vanishes due to the Seesaw relation, expressed in Eq.~(\ref{eq:unitarity_4}).

Diagrams~(f, g, h) in both, Figure~\ref{fig:gamma_diagrams} and Figure~\ref{fig:z_diagrams} vanish when the three diagrams are summed. It should also be noted that in the limit when the four-momentum squared of the outer boson is zero, $Q^2\to 0$, the sum of the three diagrams vanishes completely. The reason for this becomes clear when the calculation is done in the $R_\xi$ gauge, where the three diagrams are completely gauge-dependent.

\subsection{Branching ratio for cLFV processes}
\label{app:Br_cLFV}

The cLFV processes that receive contributions from ``HNL penguins'' or ``HNL box diagrams'' include $\ell_\beta\to \ell_\alpha\ell_\gamma\bar{\ell}_\eta$. The formulas of the branching ratios are different depending on whether $\alpha=\gamma=\eta, \alpha=\eta\neq\gamma$, or $\alpha=\gamma\neq\eta$. In the first case, there is an additional factor coming from identical fermions in the final state that is not found in the second case; and in the third case, only box diagrams contribute, $Z$ and $\gamma$ penguins do not contribute at 1-loop.

The formulas of the branching ratios for these decays are:
\begin{align}
  \mathrm{BR}(\ell_\beta\to 3\ell_\alpha) =&
  \frac{\alpha_w^4}{24576\,\pi^3}\,\frac{m_{\beta}^4}{M_W^4}\,
  \frac{m_{\beta}}{\Gamma_{\beta}}\left\{2\left|\frac{1}{2}F_\text{box}^{\beta
    3\alpha} +F_Z^{\beta\alpha} - 2 s_w^2\,(F_Z^{\beta\alpha} -
  F_\gamma^{\beta\alpha})\right|^2 \right.  \nonumber\\ 
  +& \left. 4 s_w^4\, |F_Z^{\beta\alpha} -
  F_\gamma^{\beta\alpha}|^2 + 16
  s_w^2\,\mathrm{Re}\left[(F_Z^{\beta\alpha} + \frac{1}{2}F_\text{box}^{\beta
    3\alpha})\,G_\gamma^{\beta \alpha
    \ast}\right]\right.\nonumber\\ 
  -&\left. 48 s_w^4\,\mathrm{Re}\left[(F_Z^{\beta\alpha} -
  F_\gamma^{\beta\alpha})\,G_\gamma^{\beta\alpha \ast}\right] + 32
  s_w^4\,|G_\gamma^{\beta\alpha}|^2\left[\log\frac{m_{\beta}^2}{m_{\alpha}^2}
  - \frac{11}{4}\right] \right\}\,, 
  \label{eq:3l} \\
  \mathrm{BR}(\ell_\beta\to \ell_\alpha\bar{\ell}_\alpha\ell_\gamma) =&
  \frac{\alpha_w^4}{24576\,\pi^3}\,\frac{m_{\beta}^4}{M_W^4}\,
  \frac{m_{\beta}}{\Gamma_{\beta}}\left\{\left|F_\text{box}^{\beta
    \gamma\alpha\alpha} +F_Z^{\beta\alpha} - 2 s_w^2\,(F_Z^{\beta\alpha} -
  F_\gamma^{\beta\alpha})\right|^2 \right.  \nonumber\\ 
  +& \left. 4 s_w^4\, |F_Z^{\beta\alpha} -
  F_\gamma^{\beta\alpha}|^2 + 8
  s_w^2\,\mathrm{Re}\left[(F_Z^{\beta\alpha} + F_\text{box}^{\beta\gamma\alpha\alpha})\,G_\gamma^{\beta \alpha
    \ast}\right]\right.\nonumber\\ 
  -&\left. 32 s_w^4\,\mathrm{Re}\left[(F_Z^{\beta\alpha} -
  F_\gamma^{\beta\alpha})\,G_\gamma^{\beta\alpha \ast}\right] + 32
  s_w^4\,|G_\gamma^{\beta\alpha}|^2\left[\log\frac{m_{\beta}^2}{m_{\alpha}^2}
  - 3\right] \right\}\,, 
  \label{eq:3l_2} \\
  \mathrm{BR}(\ell_\beta\to\ell_\alpha\ell_\alpha\bar{\ell}_\gamma) =&
  \frac{\alpha_w^4}{49152\,\pi^3}\,\frac{m_{\beta}^4}{M_W^4}\,
  \frac{m_{\beta}}{\Gamma_{\beta}}\left|F_\text{box}^{\beta\alpha\alpha\gamma}\right|^2
  \,. \label{eq:3l_3}
\end{align}

Another process of interest, $\mu\to e$ conversion on the nucleus $\mathrm{A}$, also receives ``penguin'' contributions. The conversion rate can be computed using the results obtained in \cite{Kitano:2002mt}. The conversion rate is:
\begin{equation}
  \label{eq:mu_e_conversion}
  \mathrm{CR}(\mu - e,\,\mathrm{A}) = \frac{2 G_F^2\,\alpha_w^2\,
    m_\mu^5}{(4\pi)^2\,\Gamma_{\text{capt.},\mathrm{A}}}
  \begin{aligned}[t]
    &\left|4 V^{(p)}_\mathrm{A}\left[s_w^2\,F_\gamma^{e\mu} + \left(\frac{1}{4} - s_w^2\right)F_Z^{e\mu} + \frac{1}{2}F_\text{box}^{\mu e uu} + \frac{1}{4}F_\text{box}^{\mu e dd}\right] \right. \\
    &\left. + 4 V^{(n)}_{\mathrm{A}}\left[-\frac{1}{4}F_Z^{e\mu} + \frac{1}{4}F_\text{box}^{\mu e uu} + \frac{1}{2}F_\text{box}^{\mu e dd} \right]  + s_w^2 \frac{G_\gamma^{\mu e}D_{\mathrm{A}}}{2 e}\right|^2\,,
  \end{aligned}  
\end{equation}
we also mention for completeness that the processes $\ell_\beta \to \ell_\alpha\gamma$ do not contain ``HNL penguins'' and
therefore do not exhibit the behavior in question:
\begin{equation}
  \label{eq:l1_l2_gamma}
  \mathrm{BR(\ell_\beta\to\ell_\alpha\gamma)} =  \frac{\alpha_w^3\,s_w^2}{256\,\pi^2}\,\frac{m_{\beta}^4}{M_W^4}\, \frac{m_{\beta}}{\Gamma_{\beta}}\, \left|G_\gamma^{\beta \alpha} \right|^2\:. 
\end{equation}

The loop functions that enter expressions~(\ref{eq:3l}--\ref{eq:l1_l2_gamma}) are given by
\begin{align}
  \label{eq:G_F_H_functions}
  G_\gamma^{\beta \alpha} =& \sum_{i=1}^{3+ \mathcal{N}} \mathcal{U}_{\alpha i}^{\phantom{\ast}}\,\mathcal{U}_{\beta i}^\ast\,
  G_{\gamma}(x_i)\,, \\
  F_\gamma^{\beta \alpha} =& \sum_{i=1}^{3+ \mathcal{N}} \mathcal{U}_{\alpha i}^{\phantom{\ast}}\,\mathcal{U}_{\beta i}^\ast\,
  F_{\gamma}(x_i)\,, \\
  F_Z^{\beta \alpha} =& \sum_{i,j = 1}^{3+ \mathcal{N}} \mathcal{U}_{\alpha i}^{\phantom{\ast}}\,\mathcal{U}_{\beta j}^\ast
  \left[\delta_{ij} F_{Z}(x_i) + \mathcal{C}^{\phantom{\ast}}_{ij} G_{Z}(x_i, x_j) + \mathcal{C}^{\ast}_{ij}\,H_{Z}(x_i,x_j)\right]\,,
\end{align}
\begin{align}
  F_\text{box}^{\beta \alpha\gamma\eta} =& \sum_{i,j = 1}^{3+ \mathcal{N}}
  \mathcal{U}_{\alpha i}^{\phantom{\ast}}\,\mathcal{U}_{\beta j}^\ast\,
  \mathcal{U}_{\gamma i}^{\phantom{\ast}}\,\mathcal{U}_{\eta j}^\ast\, G_\text{box}(x_i, x_j) - \mathcal{U}_{\beta j}^\ast \,\mathcal{U}_{\eta i}^{\ast}\left[\mathcal{U}_{\alpha i}^{\phantom{\ast}}\,\mathcal{U}_{\gamma j}^{\phantom{\ast}} + \mathcal{U}_{\gamma i}^{\phantom{\ast}}\,\mathcal{U}_{\alpha i}^{\phantom{\ast}}\right]\, F_\text{Xbox}(x_i, x_j)\,,\\
  F_\text{box}^{\mu e uu} \simeq & \sum_{i = 1}^{3+ \mathcal{N}} 
  \mathcal{U}_{e i}^{\phantom{\ast}}\,\mathcal{U}_{\mu j}^\ast F_\text{box}(x_i,0)\,, \\
  \label{eq:G_F_H_functions_1}
  F_\text{box}^{\mu e dd} \simeq & \sum_{i = 1}^{3+ \mathcal{N}} 
  \mathcal{U}_{e i}^{\phantom{\ast}}\,\mathcal{U}_{\mu j}^\ast F_\text{Xbox}(x_i,0)\,,
\end{align}
where $x_i = M_i^2/M_W^2$, $s_w^2 =\sin^2\theta_W$ is the sine squared of the Weinberg angle, $\alpha_w = g^2/4\pi$, where $g$ is the weak coupling constant, $e = \sqrt{4\pi\alpha}$ is the value of the elementary charge, and $D_\mathrm{A}, V^{(p)}_{\mathrm{A}}$, and $V^{(n)}_{\mathrm{A}}$ are all constants which depend on the nucleus in question, their precise definitions and values for difference nuclei can be found in \cite{Kitano:2002mt}. 

Equations ~(\ref{eq:G_F_H_functions}--\ref{eq:G_F_H_functions_1}) can be rewritten in terms of the light-heavy mixing angle, $\Theta$ and HNL masses alone, using the fact that $\mathcal{U}\,\mathcal{U}^\dagger = \mathbbm{1}$ and that neutrino masses are completely negligible in the limits we're considering:

\begin{align}
  \label{eq:loop_functions_start}
  G_\gamma^{\beta \alpha} =& \sum_{I=1}^{\mathcal{N}} \Theta_{\alpha I}^{\phantom{\ast}}\,\Theta_{\beta I}^\ast\,
  G_{\gamma}(x_I)\,, \\
  F_\gamma^{\beta \alpha} =& \sum_{I=1}^{\mathcal{N}} \Theta_{\alpha I}^{\phantom{\ast}}\,\Theta_{\beta I}^\ast\,
  F_{\gamma}(x_I)\,,\\
  F_Z^{\beta \alpha} =& \sum_{I,J = 1}^{\mathcal{N}} \Theta_{\alpha I}^{\phantom{\ast}}\,\Theta_{\beta J}^\ast 
  \begin{aligned}[t]
    &\left\{\delta_{IJ}\left[F_{Z}(x_I) + 2\,G_{Z}(0,x_I)\right] \right. \\
    &\left. +\,\mathcal{C}^{\phantom{\ast}}_{IJ}\left[G_{Z}(x_I, x_J) - G_{Z}(x_I, 0) - G_{Z}(0,x_J) \right] + \mathcal{C}^{\ast}_{IJ}\,H_{Z}(x_I,x_J)\right\}
  \end{aligned} \label{eq:F_Z}\\
  F_\text{box}^{\beta\alpha\gamma\eta} =& \sum_{I,J = 1}^{\mathcal{N}}
  \begin{aligned}[t]
    &-\left(\Theta_{\alpha I}\,\Theta_{\beta J}^\ast\,\delta_{\gamma\eta} + \Theta_{\gamma I}\,\Theta_{\beta J}^\ast\,\delta_{\alpha\eta}\right) \delta_{IJ}\left[F_\text{Xbox}(x_I, 0) - F_\text{Xbox}(0, 0)\right] \\  
    &-  \,\Theta_{\eta I}^{\phantom{\ast}}\,\Theta_{\beta J}^{\ast}\left(\Theta_{\alpha I} \Theta_{\gamma J} + \Theta_{\gamma I}\Theta_{\alpha J}\right)\left[F_\text{Xbox}(x_I, x_J) - F_\text{Xbox}(x_I, 0) - \right .\\ &\left .F_\text{Xbox}(0, x_J) + F_\text{Xbox}(0, 0) \right] 
     + \Theta_{\alpha I}^{\phantom{\ast}}\,\Theta_{\beta J}^{\ast}\,\Theta_{\gamma I}^{\phantom{\ast}}\,\Theta_{\eta J}^{\ast}\, G_\text{box}(x_I, x_J)\,,
  \end{aligned} \label{eq:F_box}\\
  F_\text{box}^{\mu e uu} = & \sum_{I = 1}^{\mathcal{N}} 
  \Theta_{e I}^{\phantom{\ast}}\,\Theta_{\mu I}^\ast\left[F_\text{box}(x_I,0) - F_\text{box}(0,0) \right] \,, \\  
  F_\text{box}^{\mu e dd} = & \sum_{I = 1}^{\mathcal{N}} 
  \Theta_{e I}^{\phantom{\ast}}\,\Theta_{\mu I}^\ast\left[F_\text{Xbox}(x_I,0) - F_\text{Xbox}(0,0) \right] \,.\label{eq:loop_functions_end}
\end{align}
The functions $G_\gamma, F_\gamma, F_Z, F_\text{Xbox}, F_\text{box}$ and $G_{\text{box}}$ are listed in the appendix~\ref{app:loop_functions} below.

Finally, another set of decays that also receives contributions from HNL penguins are $Z\to \ell_\alpha\bar{\ell}_\beta$ and $H\to\ell_\alpha\bar{\ell}_\beta$ \cite{Korner:1992an, Ilakovac:1994kj,Illana:1999ww, Korner:1992zk, Pilaftsis:1992st, Arganda:2004bz, Thao:2017qtn}, but neither of their constraints are competitive with the other processes mentioned. Moreover, Higgs decays to two leptons are helicity suppressed: their branching ratio is proportional to the masses of the outgoing particles, adding another suppression to it.

\subsection{Relevant combinations of the mixing angles}
\label{app:Theta_combinations}

We see that the relevant observables depend on specific combinations of products of the elements of $\Theta$.
Specifically, the branching ratios are proportional to the parameter $\Lambda_{\alpha \beta}$, defined by Eq.~2 in the main text).
HNL penguins depend on the combinations $\left|\sum\limits_{I,J} \Theta_{\alpha I}^{\phantom{\ast}} \Theta_{\beta J}^\ast\,\mathcal{C}_{IJ}\right|^2$.
In the case of two HNLs with equal masses and $\Im(\omega) \gg 1$, this expression simplifies to
\begin{equation}
  \label{eq:C_IJ}
  \left|\sum_{I,J = 1}^{\mathcal{N}} \Theta_{\alpha I}^{\phantom{\ast}} \Theta_{\beta J}^\ast\,\mathcal{C}_{IJ}\right|^2
  \simeq \Lambda_{\alpha\beta}\,U_{\mathrm{tot}}^8\,,
\end{equation}
It should be noted that there is another kind of HNL penguin diagrams, where each HNL, running in the loop, violates the \emph{total} lepton number by $+1$ or $-1$.
The resulting process is still lepton-number conserving but is proportional to the sum~(\ref{eq:C_IJ}) albeit with $\mathcal{C}^{\ast}_{IJ}$  instead of $\mathcal{C}_{IJ}$.
In this case, the sum automatically vanishes.

Another noteworthy combination involves a sum that uses the same flavor thrice, relevant for $\ell_\alpha \to 3\ell_\beta$ decays.
It can be easily evaluated using the same limits as we used to evaluate the last one:
\begin{equation}
  \left|\sum\limits_{I,J = 1}^\mathcal{N} \Theta_{\alpha I}^{\phantom{\ast}} \Theta_{\beta J}^\ast \,\Theta_{\beta I}^\ast\,\Theta_{\beta J}^{\phantom{\ast}} \right|^2
  = \lambda_\beta^2\,\Lambda_{\alpha\beta}\,U_{\mathrm{tot}}^8\,,
\end{equation}
where $\lambda_\beta = U_\beta^2/U_\mathrm{tot}^2$, with $U_\beta^2$ defined in Eq.~(\ref{eq:1}).
Moreover, considering the same case, the sum $\sum_{I,J} \Theta_{\alpha I}^{\phantom{\ast}} \Theta_{\beta J}^\ast \,\Theta_{\beta I}^{\phantom{\ast}}\,\Theta_{\beta J}^{\ast}$ would immediately vanish. 

\begin{figure}
    \centering
    \includegraphics[width=\linewidth]{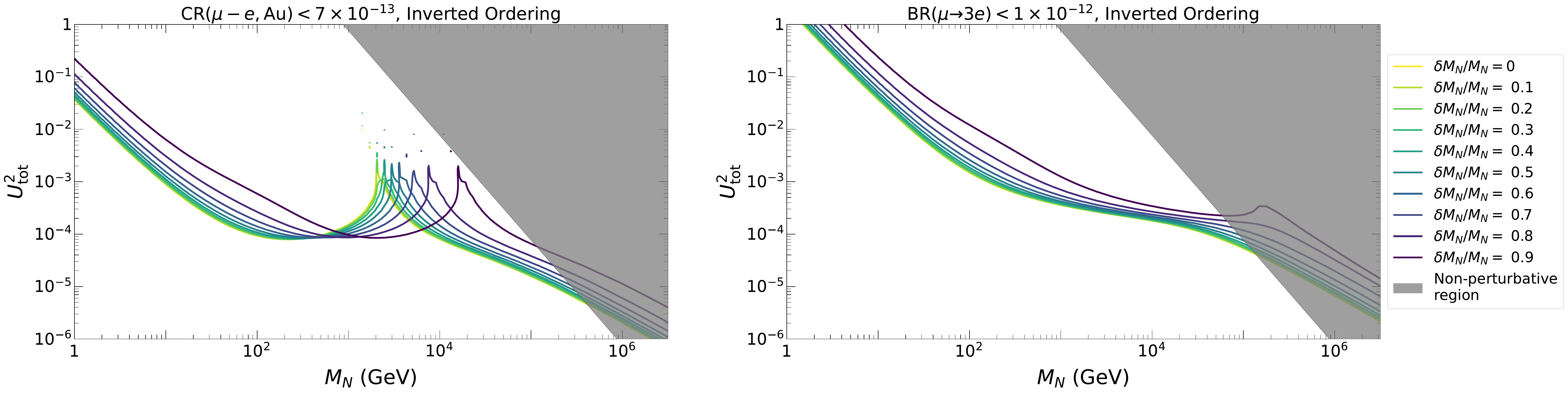}
    \caption{Constraints for a degenerate and hierarchical HNL mass spectrum for two different processes (showed at the top of the plot): muon conversion in Gold (left) and $\mu\to 3e$ (right).}
    \label{fig:mass_hierarchy_effects}
\end{figure}

These results are valid for any hierarchy of HNL masses. But it can be seen from Eq.~(\ref{eq:F_Z}) and Eq.~(\ref{eq:F_box}) that the sums are also proportional to HNL masses. In the case of degenerate HNL masses, we can take all the mass-dependent parts out of the summation, and we would only have to deal with the sum of mixing angles. If HNL masses were not degenerate, then the LNV parts would come to contribute. Figure~(\ref{fig:mass_hierarchy_effects}) shows how much the constraints would change if we had hierarchical HNLs. 

At this point, we have to stress that including hierarchical HNLs no longer protects the smallness of active neutrino masses, as there would no longer be a lepton number conserving symmetry guaranteeing their smallness. Indeed, if HNLs were to be highly hierarchical, loop contributions to neutrino masses would become larger compared to experimental results, rendering the theory unnatural \cite{Branco:1988ex,Pilaftsis:1991ug, Shaposhnikov:2006nn, Kersten:2007vk, Yu:2020gre}.

\subsection{Loop functions}
\label{app:loop_functions}

For completeness we also show the relevant loop integrals that appear in the expression~(\ref{eq:loop_functions_start}--\ref{eq:loop_functions_end}):
\begin{align}
  F_\gamma(x) =& \frac{7 x^3 - x^2 - 12x}{12(1-x)^3} - \frac{x^4 -
    10x^3 + 12x^2}{6(1-x)^4}\log x\,, \\
  G_\gamma(x) =& -\frac{x(2x^2 + 5x - 1)}{4(1-x)^3} -
  \frac{3x^3}{2(1-x)^4}\log x\,, \\
  F_Z(x) =& -\frac{5 x}{2(1 - x)} - \frac{5x^2}{2(1-x)^2}\log x\,, \\
  G_Z(x,y) =& -\frac{1}{2(x-y)}\left[\frac{x^2(1-y)}{1-x}\log x -
  \frac{y^2(1-x)}{1-y}\log y \right]\,,\\
  H_Z(x,y) =& \frac{\sqrt{xy}}{4(x-y)}\left[\frac{x^2 - 4x}{1 -
    x}\log x - \frac{y^2 - 4y}{1 - y}\log
  y\ \right]\,, \\ 
  F_\text{box}(x,y) =& 
  \begin{aligned}[t]
    &\frac{1}{x-y}\left\{\left(4 + \frac{xy}{4}\right)\left[\frac{1}{1-x} + \frac{x^2}{(1-x)^2} \log x - \frac{1}{1-y} - \frac{y^2}{(1-y)^2}\log y\right]\right. \\ 
    &\left. -2xy\left[\frac{1}{1-x} + \frac{x}{(1-x)^2}
    \log x - \frac{1}{1-y} - \frac{y}{(1-y)^2}\log y
    \right]\right\}\,
  \end{aligned} \\
  F_\text{Xbox}(x,y) =& 
  \begin{aligned}[t]
    &-\frac{1}{x-y}\left\{\left(1 + \frac{xy}{4}\right)\left[\frac{1}{1-x} + \frac{x^2}{(1-x)^2} \log x -\frac{1}{1-y} - \frac{y^2}{(1-y)^2}\log y\right]\right. \\
    &\left. -2xy\left[\frac{1}{1-x} + \frac{x}{(1-x)^2}
    \log x - \frac{1}{1-y} - \frac{y}{(1-y)^2}\log y
    \right]\right\}\,,  
  \end{aligned} \\
  G_\text{box}(x,y) =& 
  \begin{aligned}[t]
    &-\frac{\sqrt{xy}}{x-y}\left\{(4 + xy)\left[\frac{1}{1-x} + \frac{x}{(1-x)^2} \log x - \frac{1}{1-y} - \frac{y}{(1-y)^2}\log y\right]\right.\\
    &\left. -2\left[\frac{1}{1-x} + \frac{x^2}{(1-x)^2}
    \log x - \frac{1}{1-y} - \frac{y^2}{(1-y)^2}\log y
    \right]\right\}\,.\\ 
  \end{aligned} 
\end{align}

\subsection{Asymptotic behaviour of branching ratios}
Using the formulas defined in Appendix~\ref{app:loop_functions}, we can extract the behaviour of branching ratios~(\ref{eq:3l}--\ref{eq:mu_e_conversion}) for large masses $x=(M_N/m_W)^2\gg1$. 

\begin{align}
  \label{eq:asymptotic_3l}
  \mathrm{BR}(\ell_\beta\to 3\ell_\alpha) &\propto 
  \begin{aligned}[t]
    &U_\mathrm{tot}^8 \biggl[x^2 \left(\frac{1}{2} - 2s_w^2 + 3s_w^4 -\frac{1}{2}\lambda_\alpha + s_w^2\lambda_\alpha + \frac{1}{8}\lambda_\alpha^2 \right) \\
    &\quad+ x\log x\,\lambda_\alpha\left(-3 + 6\,s_w^2 + \frac{3}{2}\lambda_\alpha \right) 
    + \frac{9}{2}(\log x)^2 \lambda_\alpha^2\biggr] \\
    &+U_\mathrm{tot}^6 \biggl[x\log x\left(3  - \frac{34}{3}s_w^2 + 16s_w^4 - \frac{3}{2}s_w^2\lambda_\alpha + \frac{8}{3}s_w^4\lambda_\alpha \right) \\
    &\quad+ (\log x)^2\,(12s_w^2 - 9)\,\lambda_\alpha\biggr]
    +U_\mathrm{tot}^4 (\log x)^2\left(\frac{9}{2} - 16s_w^2 + \frac{64}{3}s_w^4 \right)\,, 
  \end{aligned} \\
  \label{eq:asymptotic_muon_conversion}
   \begin{split}
  \mathrm{CR}(\mu - e,\,\mathrm{A}) &\propto \Bigl|U_\mathrm{tot}^4\,x\,(3- 8s_w^2)\left(V^{(n)}_\mathrm{A} - V^{(p)}_\mathrm{A}\right) \\
  &\quad+ U_\mathrm{tot}^2\log x\left[V^{(n)}_\mathrm{A}\left(9 - 24s_w^2 \right) - V^{(p)}_\mathrm{A}\left(9 - 20s_w^2\right) \right]\Bigr|^2\,,
  \end{split}
\end{align}
We have only considered the leading terms in $x$ for each power of ${U}_\mathrm{tot}$ in equations~(\ref{eq:asymptotic_3l}--\ref{eq:asymptotic_muon_conversion}), there are additional powers of $x$ that are not negligible and should be taken into consideration when doing any calculation regarding them.

Equations~(\ref{eq:3l_2}--\ref{eq:3l_3}) also show a similar asymptotic behaviour to that of~(\ref{eq:3l}), as they also have contributions coming from $Z$-penguins and HNL box diagrams. The decays these formulas describe ($\tau\to\mu\mu e,\tau\to ee\mu$) are not as constrained as muon decays and were not examined, but they do offer an interesting opportunity to probe all mixing angles since these decays require HNLs to couple to all lepton flavors.

\section{Feynman diagrams contributing to the relevant processes} \label{app:diagrams}
We show below all the relevant Feynman diagrams of the processes we highlighted. All are in the Feynman-'t Hooft gauge and therefore contain intermediate Goldstone bosons.
The diagrams containing divergences have their respective letter in red, and the ones that exhibit the non-decoupling behavior highlighted in the main text are in green.
\begin{figure}[H]
    \centering
    \includegraphics[width = \linewidth]{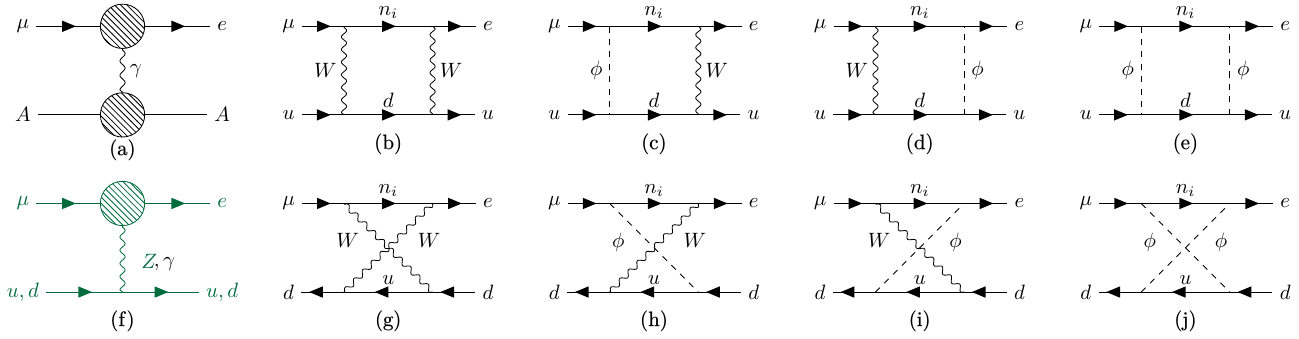}
    \caption{Diagrams contributing to muon conversion in a nucleus decay. 
    The diagrams in green are new, accounting for two-HNL interactions.
    Notice that in this case all two HNL contributions are inside the ``blob'' of the diagram (f) coupling to $Z$-bosons (given by  the diagram~\ref{fig:z_diagrams}), which explains its color.
    }
    \label{fig:muon_conversion}
\end{figure}

\section{Higher order terms} \label{app:higher_order_terms}
\begin{figure}[H]
    \centering
    \includegraphics{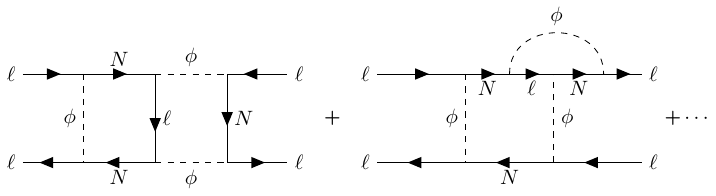}
    \caption{Example of two-loop diagrams contributing to $\ell_\beta \to 3 \ell_\alpha$}
    \label{fig:two_loops}
\end{figure}
A complete computation of all two-loop diagrams of all non-decoupling processs is beyond the scope of this work, but by looking at the parameter dependence of the amplitude of the leading two-loop non-decoupling diagrams, one can get an idea of how the calculation will be validity of the supposition by keeping $Y < 4\pi$, the theory remains perturbative. 

Indeed, by finding the parametric dependence of both diagrams in Fig.~\ref{fig:two_loops} we find that:
\begin{equation}
    \mathcal{M}_{\mathrm{2-loop}} = \Theta^2 \left(\frac{Y^2}{16\pi^2}\right)^2\,,
\end{equation}

where the $16\pi^2$ is the usual term one gets when doing the four dimensional loop integrals, which in case of a two-loop diagram it should be squared, of course. This is in direct comparison with the parametric dependence of the leading order terms of one-loop diagrams:

\begin{equation}
    \mathcal{M} = \Theta^2\left[\frac{Y^2}{16\pi^2} + \left(\frac{Y^2}{16\pi^2}\right)^2 \right]\,,
\end{equation}

The two-loop amplitude will be smaller than the one-loop one as long as $Y < 4\pi$, which is in complete accordance with the classical perturbative limit for Yukawa couplings. One should expect the higher order ones to follow a similar dependence on Yukawas and a suppression of $16\pi^2$.

\bibliographystyle{JHEP}
\bibliography{bibliography.bib}
\end{document}